\documentclass[12pt]{article}
\usepackage{amsmath}
\usepackage{epsf}
\usepackage{graphicx}
\usepackage{amsfonts}   
\usepackage{amssymb}
\usepackage{cite} 
\newcommand{\be}{\begin{equation}}
\newcommand{\ee}{\end{equation}}
\newcommand{\beq}{\begin{equation}}
\newcommand{\eeq}{\end{equation}}
\newcommand{\bea}{\begin{eqnarray}}
\newcommand{\eea}{\end{eqnarray}}
\newcommand{\beqa}{\begin{eqnarray}}
\newcommand{\eeqa}{\end{eqnarray}}
\newcommand{\ba}{\begin{array}}
\newcommand{\ea}{\end{array}}

\newcommand{\cO}{{\cal O}}
\newcommand{\cB}{{\cal B}}

\newcommand{\no}{\nonumber}

\newcommand{\lsim}{\stackrel{<}{_\sim}}
\newcommand{\gsim}{\stackrel{>}{_\sim}}

\newcommand{\diag}{\mathrm{diag}}
\newcommand{\hc}{{\rm h.c.}}
\newcommand{\RE}{{\rm Re}}
\newcommand{\IM}{{\rm Im}}

\newcommand{\kppn}{K^+\rightarrow\pi^+\nu\bar\nu}
\newcommand{\klpn}{K_{\rm L}\rightarrow\pi^0\nu\bar\nu}
\newcommand{\kpnn}{K \rightarrow\pi\nu\bar\nu}

\newcommand{\kLpll}{K_L \rightarrow\pi^0 \ell^+\ell^-}
\newcommand{\kLpee}{K_L \rightarrow\pi^0 e^+e^-}
\newcommand{\kLpmm}{K_L \rightarrow\pi^0 \mu^+\mu^-}

\newcommand{\sttop}{{\tilde t}}

\newcommand{\charg}{{\tilde \chi}}

\newcommand{\AU}{\mathbf{A}_{U}}
\newcommand{\AD}{\mathbf{A}_{D}}
\newcommand{\MU}{\mathbf{M}_{U}}
\newcommand{\MD}{\mathbf{M}_{D}}
\newcommand{\MQ}{\mathbf{M}_{Q}}
\newcommand{\Muu}{\mathbf{\hat M}_{\tilde{u}}}
\newcommand{\Mdd}{\mathbf{\hat M}_{\tilde{d}}}
\newcommand{\MMuu}{\mathbf{M}_{\tilde{u}}}
\newcommand{\MMdd}{\mathbf{M}_{\tilde{d}}}
\newcommand{\MuuLL}{\mathbf{\hat M}_{\tilde{u} LL}}
\newcommand{\MuuRL}{\mathbf{\hat M}_{\tilde{u} LR}}
\newcommand{\MuuRR}{\mathbf{\hat M}_{\tilde{u} RR}}
\newcommand{\MMuuAA}{\mathbf{M}_{\tilde{u} AA}}
\newcommand{\MMuuBB}{\mathbf{M}_{\tilde{u} BB}}
\newcommand{\MMuuAB}{\mathbf{M}_{\tilde{u} AB}}
\newcommand{\MddLL}{\mathbf{\hat M}_{\tilde{d} LL}}
\newcommand{\MddRL}{\mathbf{\hat M}_{\tilde{d} LR}}
\newcommand{\MddRR}{\mathbf{\hat M}_{\tilde{d} RR}}

\begin{document}

\thispagestyle{empty}
\setcounter{page}{0}
\begin{flushright}
April 2006 \\
SFB/CPP-06-16 \\
TTP06-13\\ 
\end{flushright}
\vskip   2 true cm 
\begin{center}
{\Large \textbf{Exploring the flavour structure of the MSSM \\[5pt] 
                 with rare $K$ decays}} \\ [20 pt]
\textsc{Gino Isidori},${}^{1}$
\textsc{Federico Mescia},${}^{1}$
\textsc{Paride Paradisi},${}^{2}$ \\[5pt] 
\textsc{Christopher Smith},${}^{3}$  
\textsc{Stephanie Trine}${}^{4}$  \\ [20 pt]
${}^{1}~$\textsl{INFN, Laboratori Nazionali di Frascati, I-00044 Frascati,
      Italy} \\ [5 pt]
${}^{2}~$\textsl{INFN, Sezione di Roma II and Dipartimento di Fisica, 
      \\ Universit\`a di Roma ``Tor Vergata'', I-00133 Rome, Italy} \\ [5 pt]
${}^{3}~$\textsl{Institut f\"ur Theoretische Physik, Universit\"at Bern, 
     CH-3012 Bern, Switzerland} \\ [5 pt]
${}^{4}~$\textsl{Institut f\"ur Theoretische Teilchenphysik, \\
      Universit\"at Karlsruhe, D-76128 Karlsruhe, Germany} \\[25pt]
\textbf{Abstract}
\end{center}
\noindent  
We present an extensive analysis of rare $K$ decays,
in particular of the two neutrino modes
$K^{+}\rightarrow\pi^{+}\nu\overline{\nu}$
and $K_L\rightarrow\pi^0\nu\overline{\nu}$, 
in the Minimal Supersymmetric extension of the Standard Model. 
We analyse the expectations for the branching ratios 
of these modes, both within the restrictive framework 
of the minimal flavour violation hypothesis and within 
a more general framework with new sources 
of flavour-symmetry breaking. In both scenarios, 
the information that can be extracted from precise measurements 
of the two neutrino modes turn out to be very useful 
in restricting the parameter space of the model,
even  after taking into account the possible 
information on the mass spectrum derived from 
high-energy colliders, and the constraints from $B$-physics experiments. 
In the presence of new sources of flavour-symmetry breaking, 
additional significant constraints on the model can be derived also from 
the two  $K_L\rightarrow\pi^0\ell^+\ell^-$ modes. 

\newpage

\section{Introduction}
As widely discussed in the literature, rare decays dominated
by one-loop electroweak dynamics offer a very powerful tool to investigate
the flavour structure of physics beyond the Standard Model (SM) \cite{Reviews}.
Among them, the processes $K^{+}\rightarrow\pi^{+}\nu\overline{\nu}$
and $K_L\rightarrow\pi^0\nu\overline{\nu}$ are certainly a privileged observatory
because of the high-level of accuracy achieved in their theoretical description
\cite{NNLO,NLO,IMS,Long_old,BI,MP}: future precise measurements of these modes
will have a non-trivial impact on physics well above the electroweak scale \cite{BBIL}.

In the last few years precise measurements of
flavour-changing neutral-current (FCNC)
processes in the $B$ sector have severely 
restricted the parameter space of new-physics
models, especially in the flavour sector. Moreover, 
a direct exploration of the physics in the TeV range 
is expected soon with the start of the LHC program. 
Within this context, it is worth 
to understand if, and at which level, 
the indirect information 
which could be extracted from rare $K$ decays 
is still useful. The main purpose 
of the present paper is an  attempt 
to answer this question, within the 
specific framework of the 
Minimal Supersymmetric extension of the SM (MSSM).

Several analyses of rare $K$ decays 
within the MSSM have already been presented in the 
literature, both in the general framework of arbitrary 
new sources of flavour mixing \cite{NW,BRS,CI,Kpnn_SUSY1,Jager,Kpnntgb},
and also in well-motivated scenarios with more restrictive 
hypotheses \cite{Kpnn_MFV1}. The main purpose of all these works 
has been the identification of the maximal deviations 
from the SM of the two $K \to \pi \nu\overline{\nu}$ rates.
Our analysis has a different goal: understanding how 
precise measurements of these observables 
can be used to discriminate among different 
versions of the MSSM. We will analyse in particular 
two general frameworks:
\begin{itemize} 
\item[I.] The most general version of the MSSM 
compatible with the Minimal Flavour Violation (MFV) 
hypothesis, as defined in Ref.~\cite{MFV}.
\item[II.] The MSSM with generic new sources of
flavour-symmetry breaking, in particular with sizable 
non-MFV trilinear soft-breaking terms 
in the up sector ($\AU$), 
with $R$-parity conservation 
and moderate values of $\tan\beta$ ($\tan\beta \lsim 30$).
\end{itemize}
As we will show, in both these frameworks precise measurements 
of the two $\cB(K \to \pi \nu\overline{\nu})$ are very 
useful to determine the (flavour) structure of the 
model. This statement remains true even taking into 
account possible future constraints on the MSSM mass 
spectrum obtained at the LHC, and the refinement of the
flavour constraints expected from $B$ factories.
Within the scenario II, we will show in particular 
that present constraints still allow a large freedom 
concerning the flavour structure of the $\AU$ terms.
In the presence of sizable deviations from the MFV hypothesis 
in this sector, a key role is played also by the 
two $K_L\rightarrow\pi^0\ell^+\ell^-$ modes. 

The paper is organized as follows: in Section~\ref{sec:basic}
we recall some basic formulae for the evaluation of 
rare $K$ decay branching ratios. In Section~\ref{sec:MFV} 
we: i) introduce the MFV scenario; ii) analyse the expectations 
of the two $\cB(K \to \pi \nu\overline{\nu})$ in this framework;
iii) discuss the consequences of these findings and 
compare them with the previous literature. 
Similarly, in Section~\ref{sec:non_MFV} we introduce
and analyse the consequences for rare $K$ decays of the  
scenario II. The main results are summarized in the Conclusions.

\section{Basic formulae for rare $K$ decays}
\label{sec:basic}

Within the class of models
considered here, the supersymmetric 
contributions to $\kpnn$ decays can be described to a good 
accuracy in terms of a single complex function\footnote{~There  
are two notable corners of the MSSM parameter space (which 
will not be analysed in this work) where 
this approximation is not valid: the large-$\tan\beta$ scenario 
with non-MFV right-right mixing terms, discussed recently in~\cite{Kpnntgb}, 
and models with large violations of lepton-flavour 
universality, discussed in~\cite{GIM2}.}
\be
W = \frac{1}{3} \sum_{l=e,\mu,\tau}  W_l~,
\ee
where $W_l$ are the Wilson coefficients of the following effective Hamiltonian: 
\be
\label{Heff}
{\cal H}_{eff}^{|\Delta S| = 1} =  \frac{G_F}{\sqrt{2}} \frac{ \lambda^5 \alpha_{\rm em} }{2 \pi \sin^2\theta_W}
\sum_{l=e,\mu,\tau} W_l~\bar s \gamma^\mu(1-\gamma_5)d
~\bar\nu_l \gamma_\mu(1-\gamma_5) \nu_l ~+~\hc~
\ee
with $\lambda = |V_{us}| = 0.225 \pm 0.001$.
In terms of this function, the two $\kpnn$ branching ratios can be written as
\bea
\cB\left(  K^{+}\rightarrow\pi^{+}\nu\bar{\nu}\right)   &=& \kappa_+
 \left|  W_{\rm SM} + W_{\rm SUSY} \right|^{2} \\
\cB\left(  K_L \rightarrow\pi^{0}\nu\bar{\nu}\right)   &=& 
  \kappa_L \left[  \IM (W_{\rm SM} + W_{\rm SUSY})\right]^{2}
\eea
where\footnote{~The numerical values of the $\kappa_i$
--which encode isospin-breaking and $SU(3)$ violations 
in the $K\to \pi$ matrix elements \cite{MP}-- have been 
updated with respect to the previous literature 
taking into account the latest results on the
$K\to \pi$ form factors from Ref.~\cite{CKM_Cabibbo}. 
The same comment applies to  the $C^\ell_i$ in Eq.~(\ref{eq:Cli}). }
$\kappa_+ = (5.26\pm 0.06) \times  10^{-11}$ and  
$\kappa_L = (2.29\pm 0.03) \times 10^{-10}$.
Defining further 
$\lambda_q = V_{qs}^* V_{qd}$
(where $V_{ij}$ denotes the generic element of the Cabibbo-Kobayashi-Maskawa matrix),
the SM contribution to the $W$ function reads 
\be
 W_{\rm SM} = \frac{\RE \lambda_{c}}{\lambda} P_{u,c} + \frac{ \RE \lambda_t}{\lambda^5} X_t
 + i \frac{ \IM \lambda_t}{\lambda^5} X_t~,
\ee
with $X_t = 1.464\pm 0.041$ \cite{NLO} and $P_{u,c} = 0.41\pm 0.04$~\cite{NNLO,IMS}.
For the numerical values of the other input parameters we refer to Refs.~\cite{CKM_Cabibbo,UTfit}.

As discussed in \cite{NW,BRS,Kpnn_SUSY1,CI,Jager},
among the additional non-standard contributions to the  
$W$ function appearing in the MSSM ($W_{\rm SUSY}$), 
only those associated with chargino up-squark loops and 
charged-Higgs top-quark loops can compete in size with $W_{\rm SM}$.
A complete listing of these contributions can be found in Ref.~\cite{Jager}.

The effective Hamiltonian necessary 
to describe the two $\kLpll$ modes can in general 
be written as  
\be
{\cal H}_{eff}^{|\Delta S| = 1} = \frac{G_F}{\sqrt{2}} \frac{\lambda^5 \alpha_{\rm em} }{2\pi} \, 
 \Big[ \, \sum_{i} w^e_i(\mu)  Q^e_i(\mu) + w^\mu_i(\mu)  Q^\mu_i(\mu)  \Big]\, +\, \mbox{h.c.}~,
\label{eq:heff_ll}
\ee
where the list of potentially relevant operators includes 
four-quark operators, photon- and gluon-dipole operators, and 
\be
Q^\ell_{7V} \, =  \, \overline{s} \gamma^{\mu}(1-\gamma_5) d \, \overline{\ell} 
\gamma_{\mu} \ell~, \qquad \quad
  Q^\ell_{7A} \, =  \, \overline{s} \gamma^{\mu} (1-\gamma_5) d \, \overline{\ell}
\gamma_{\mu} \gamma_5 \ell~.
\ee 
Both within the SM and in the class
of supersymmetric models we are considering, 
the direct-CP-violating transition 
$K_2 \to \pi^0 \ell^+\ell^-$ turns out to be 
dominated by the contributions of $Q^\ell_{7V,A}$.
In this limit, the corresponding $K_L$ branching 
ratios can be written as \cite{BDI}
\be
\cB(K_{L}\rightarrow\pi^{0}\ell^{+}\ell^{-})=\left( C_{\rm mix}^{\ell}
+ C_{\rm int}^{\ell} + C_{\rm dir}^{\ell} + C_{\rm CPC}^{\ell}\right) \times10^{-12},
\ee
with the following set of coefficients:
%
%
%
\be
\begin{array}[c]{ll}
   C_{\rm int}^{e} =   - (7.47 \pm 0.20) \times |a_{S}| \times \IM~w^e_{7V}~,  
&  C_{\rm mix}^{e} =   (13.9 \pm 0.5) \times |a_{S}|^{2}~, \\
   C_{\rm int}^{\mu} = - (1.77 \pm 0.04) \times |a_{S}| \times \IM~w^\mu_{7V} ~, 
&  C_{\rm mix}^{\mu} = (3.27  \pm 0.2) \times |a_{S}|^{2}~, \\
   C_{\rm dir}^{e}   = (2.02  \pm 0.10)  \times \left[  (\IM~w^e_{7V})^2 + (\IM~w^e_{7A})^2 \right]~, 
&  C_{\rm CPC}^{e} \approx 0~, \\
   C_{\rm dir}^{\mu}  = (1.12 \pm 0.05) \times \left[0.43 \times (\IM~w^\mu_{7V})^2 + (\IM~w^\mu_{7A})^2 \right]~,
&  C_{\rm CPC}^{\mu}  = (5.2\pm 1.6)~, \\
\end{array}
\label{eq:Cli}
\ee
Here $|a_{S}|= 1.2 \pm 0.2$ denotes the non-perturbative low-energy 
constant extracted from the experimental value of 
$\cB(K_{S}\rightarrow\pi^{0}\ell^{+}\ell^{-})$ \cite{NA48},
and the Wilson coefficients are renormalized at the scale 
$\mu_{\rm IR} \approx 1$~GeV. Following the analyses of \cite{BDI,EdR}, we have  
assumed a positive interference between the long-distance 
amplitude and the SM short-distance contribution.
Using the notation of \cite{BLMM}, the latter can be expressed as
\bea
\left( \IM w^\ell_{7V} \right)_{\rm SM} &=&  - y_{7V} \times
\frac{2\pi \IM \lambda_t}{\lambda^5 \alpha_{\rm em} } ~=~ - (0.73 \pm 0.04) 
\times \frac{2\pi  \IM \lambda_t}{\lambda^5}~,  \\
\left( \IM w^\ell_{7A} \right)_{\rm SM} &=&  - y_{7A} \times 
\frac{2\pi \IM \lambda_t}{\lambda^5\alpha_{\rm em} } ~=~ (0.68 \pm 0.03)\times
\frac{2\pi \IM \lambda_t}{\lambda^5}~.
\eea
The complete analytic expressions of the chargino up-squark loop
contributions to $w^\ell_{7V,A}$ can be found in \cite{SUSY_bsll}.

\section{The MFV framework}
\label{sec:MFV}

\subsection{Definition of the model}
The MSSM with $R$ parity conservation and generic supersymmetry 
soft-breaking terms has a huge number of free parameters.
One of the virtues of $\cB(\kpnn)$ is that these 
observables are sensitive only 
to a limited subset of such parameters: those appearing 
in the chargino and up-squark mass matrices, and those which 
determine the charged-Higgs mass.
To define the structure of the MSSM we are considering,
it is therefore sufficient to specify the value of 
these mass terms.

The chargino mass matrix  in the basis of electroweak 
eigenstates (wino and higgsino) is
\beq 
M_\chi = \left( \begin{array}{cc} M_2
& \sqrt{2} M_W \sin\beta \\ \sqrt{2} M_W \cos\beta & \mu \end{array}
\right)~, 
\label{eq:M_chi}
\eeq 
where the first index of both rows and columns refers to the
wino state. Here $\mu$ denotes the supersymmetric Higgs 
bilinear coupling, $M_2$ the soft supersym\-me\-try--breaking wino mass
and $\tan\beta = v_{u}/v_{d}$ the ratio of the two Higgs vacuum 
expectation values.  Note that $M_2$ can always be chosen real, without 
loss of generality.

Two constraints on the free parameters of 
$M_\chi$ could be obtained by measuring chargino masses,
while independent information could be extracted from
the cross sections of various electroweak processes.
It is therefore reasonable to assume that the complete 
structure of $M_\chi$ will be determined, to a good extent,  
by high-energy experiments.\footnote{~An additional significant 
information is obtained by flavour-conserving low-energy 
experiments, in particular by the electric-dipole-moments (e.d.m.) of quarks and 
leptons, which already provide stringent constraints 
on the possible CP violating phase of $\mu$ \cite{Pokorski}.} 
The situation is very different in the squark 
sector, where the large number of free parameters does
not allow a model-independent extraction in terms
of high-energy data only.

The soft-breaking terms appearing in the squark sector 
are the $3\times 3$ matrices $\MQ^2$,  $\MU^2$,  
$\MD^2$ (bilinear terms), and $\AU$, $\AD$ (trilinear terms).
Performing in the squark sector the same unitary 
rotations which allow to diagonalize quark mass 
matrices,
\be
\tilde{\mathbf{q}}_{L,R} ~\to~ \mathbf{V}_{L,R}^{q} \tilde{\mathbf{q}}_{L,R} \qquad 
V_{\rm CKM} =\mathbf{V}_{L}^{u}\mathbf{V}_{L}^{d\dagger}
\ee
(i.e.~adopting the so-called super-CKM basis),
the $6\times 6$ squark mass matrices assume the form 
\be
\MMuu^{2} ~\to~
\Muu^{2} = \left(
\begin{array}[c]{cc}
  \MuuLL^2 &   \MuuRL^{2\dagger} \\
  \MuuRL^2 &   \MuuRR^2
\end{array} \right)~,  \qquad 
\MMdd^{2} ~\to~
\Mdd^{2} = \left(
\begin{array}[c]{cc}
  \MddLL^2 &   \MddRL^{2\dagger} \\
  \MddRL^2 &   \MddRR^2
\end{array} \right)~, 
\label{eq:MUMD}
\ee
where
\bea
 \MuuLL^2
 &=&\mathbf{V}_{L}^{u}\MQ^{2}\mathbf{V}_{L}^{u\dagger}
+\mathbf{m}_{u}^{2}+\mathbf{1}\frac{1}{6}(4M_{W}^{2}-M_{Z}^{2})\cos2\beta ~, \\
 \MuuRR^2
 &=&\mathbf{V}_{R}^{u}\MU^{2}\mathbf{V}_{R}^{u\dagger}
+\mathbf{m}_{u}^{2}+\mathbf{1}\frac{2}{3}M_{Z}^{2}\cos2\beta\sin^{2}\theta_W~,\\
 \MuuRL^2
 &=&v_{u}\mathbf{V}_{R}^{u}\AU\mathbf{V}_{L}^{u\dagger}
-\cot\beta\mu^{\ast}\mathbf{m}_{u} ~,
\eea
with $\mathbf{m}_{u}=\diag(m_u,m_c,m_t)$, 
and similarly for the down sector. For later convenience, we also define the couplings
\be
\left(\delta_{AB}^{U}\right)_{ij} = \frac{ (  \MMuuAB^2 )_{ij} 
}{  \vert (  \MMuuAA^2 )_{ii} \vert^{1/2}
\vert (  \MMuuBB^2 )_{jj} \vert^{1/2} }
\qquad 
(A,B = L,R)
\label{eq:delta}
\ee
which parametrize the amount of flavour-symmetry breaking in 
the up-squark sector (in a generic squark basis).

The flavour structure both of the SM and the MSSM 
is characterized by a global $SU(3)^5$ flavour symmetry,
broken only by mass terms and Yukawa interactions \cite{MFV}.
Within the SM, the Yukawa interaction is the only 
source of $SU(3)^5$ breaking. Within the MSSM, there 
are in general several new flavour-symmetry 
breaking sources, encoded in the soft-breaking 
terms. In the following we shall concentrate on the 
MFV scenario, which can be considered as the most 
restrictive assumption about the structure of 
these additional flavour-symmetry breaking terms. 

According to the MFV hypothesis, the SM Yukawa matrices 
are the only source of breaking of the 
$SU(3)^5$ flavour symmetry also beyond the SM \cite{MFV}.
Neglecting suppressed terms (proportional to 
high powers of off-diagonal CKM terms and/or 
light quark masses), this symmetry 
principle implies the following structure for
the soft-breaking terms of the quark sector 
(in a generic basis for the electroweak eigenstates) \cite{MFV}:
\bea
\MQ^{2}  &=& \tilde{m}^{2}\left(  \tilde{a}_{1}\mathbf{1}
 +\tilde{b}_{1}\mathbf{Y}_{u}^{\dagger}\mathbf{Y}_{u}
 +\tilde{b}_{2}\mathbf{Y}_{d}^{\dagger}\mathbf{Y}_{d}
 +\tilde{b}_{3}\left(  \mathbf{Y}_{d}^{\dagger}\mathbf{Y}_{d}\mathbf{Y}_{u}^{\dagger}\mathbf{Y}_{u}
   +\mathbf{Y}_{u}^{\dagger}\mathbf{Y}_{u}\mathbf{Y}_{d}^{\dagger}\mathbf{Y}_{d}\right) \right) 
\no\\
\MU^{2}  &=& \tilde{m}^{2}\left(  \tilde{a}_{2}\mathbf{1}
  +\tilde{b}_{4}\mathbf{Y}_{u}\mathbf{Y}_{u}^{\dagger}\right)~, 
\qquad \mathbf{A}_{U}=A\mathbf{Y}_{u}\left( \tilde{a}_{4}\mathbf{1}
  +\tilde{b}_{6}\mathbf{Y}_{d}^{\dagger}\mathbf{Y}_{d}\right) 
\no\\
\MD^{2}  &=& \tilde{m}^{2}\left(  \tilde{a}_{3}\mathbf{1}
  +\tilde{b}_{5}\mathbf{Y}_{d}\mathbf{Y}_{d}^{\dagger}\right)~, \qquad 
\AD=A\mathbf{Y}_{d}\left(  \tilde{a}_{5}\mathbf{1}
  +\tilde{b}_{7}\mathbf{Y}_{u}^{\dagger}\mathbf{Y}_{u}\right)~,
\eea
where $\tilde{a}_{i}$ and $\tilde{b}_{i}$ are free $\cO(1)$
parameters. Redefining these parameters 
absorbing the dimensional factors $\tilde{m}$ and $A$,
this implies for the up squark mass matrix ($V \equiv V_{\rm CKM}$)
\bea
\MuuLL^2  &=&  
  a_{1}^{2}\mathbf{1}+\frac{b_{1}^{2}}{v_{u}^{2}} \mathbf{m}_{u}^{2} 
  +\frac{b_{2}^{2}}{v_{d}^{2}} V \mathbf{m}_{d}^{2} V^\dagger
+\frac{b_{3}^{2}}{v_{u}^{2}v_{d}^{2}}
  \left( V \mathbf{m}_{d}^{2}V^{\dagger}\mathbf{m}_{u}^{2}
  + \mathbf{m}_{u}^{2}V\mathbf{m}_{d}^{2} V^\dagger \right) \no \\
&& +\mathbf{m}_{u}^{2}+\mathbf{1}\frac{1}{6}(4M_{W}^{2}-M_{Z}^{2})\cos2\beta ~, 
  \no \\
\MuuRR^2  &=& a_{2}^{2}\mathbf{1}+\frac{b_{4}^{2}}{v_{u}^{2}}
  \mathbf{m}_{u}^{2} +\mathbf{m}_{u}^{2}+\mathbf{1}\frac{2}{3}M_{Z}^{2}\cos2\beta\sin^{2}\theta_{W}~,\no \\
\MuuRL^2 &=& a_{4}\mathbf{m}_{u}
  +\frac{b_{6}}{v_{d}^{2}}\mathbf{m}_{u}V\mathbf{m}_{d}^{2}V^{\dagger}
-\cot\beta\mu^{\ast}\mathbf{m}_{u} ~,
\label{eq:MFV_par}
\eea
in the super-CKM basis. Note that in principle, 
$a_4$ and $b_6$ can be complex.

\subsection{Scanning of the parameter space} 
Employing the parametrizations in Eqs.~(\ref{eq:MFV_par})
and (\ref{eq:M_chi}) we have performed a 
systematic scan of the free parameters of the model
and analysed  the consequences for the two $\cB(\kpnn)$.
Within the MFV framework, the deviations with respect to the 
SM are experimentally undetectable in
a good fraction of the parameter space. For this reason,
we concentrated in particular to identify under which 
conditions (within this restricted scenario)
it is possible to generate sizable (detectable) 
enhancements with respect to the SM in the two $\cB(\kpnn)$.
The free parameters have been varied in a wide range, 
checking the consistency with tree-level vacuum stability bounds \cite{vacuum},
direct experimental constraints on squark 
and gaugino masses (see Table~\ref{tab:expM})
and existing constraints from precision measurements 
(both in the electroweak and in the flavour sector).
The scan has been performed using the adaptive 
integration routine VEGAS \cite{VEGAS}, to search
for maximal effects and to deal most effectively 
with invalid regions in parameter space. 

\begin{table}[t]
\begin{center}
\begin{tabular}{||rl||l|l||}
\hline
\multicolumn{2}{||c||}{Ranges  (GeV) } & 
\multicolumn{2}{c||}{Exp. bounds (GeV) }  \\  \hline
     $a_{1-3}$ : &$[0,~1000]$ &
$M_{\chi_{1,2}^{\pm}}> 94$  &   $M_{\tilde{c},\tilde{u},\tilde{s},\tilde {d}}>250$ 
\\
     $|a_{4}|$ : &$[0,~3000]$ &
$M_{\chi_{1}^{0}}~(M_{\chi_{2}^{0}})  >  46~(63)$   &   $M_{\tilde{t}}>96$ 
\\    
     $|\mu|$ : &$[0,~~500] $  
& $M_{\chi_{3}^{0}}~(M_{\chi_{4}^{0}})  >  100~(116)$ &   $M_{\tilde{b}}>89$ 
\\
  $M_{2}$ : &$[0,~3000]$  & &
\\
\hline
\end{tabular}
\end{center}
\caption{ \label{tab:expM}
Ranges adopted in the scan for the free parameters of the MFV framework 
relevant for $K\rightarrow\pi\nu\overline{\nu}$.
Experimental lower bounds on squark and gaugino masses~\cite{PDG}.}
\end{table}

The chosen ranges of the relevant 
parameters are reported in Table~\ref{tab:expM}.
The parameter $a_{5}$ and all the $b_i$ have a negligible effect on
the considered branching ratios and can be set to zero 
without loss of generality. The MFV implementation of 
vacuum stability bounds then assumes the simple form 
$|a_4|^2 \leq 3(a_1^2+a_2^2)$. Also $a_{3}$ has a very small impact,
but has to be taken sufficiently large to generate down 
squark masses large enough to pass the experimental bounds.
A similar comment applies to the gaugino mass parameter 
$M_{1}$ in connection with the experimental bounds on neutralino 
masses. For definiteness, we have taken $M_1=500$~GeV.
As far as $\tan\beta$ is concerned, we have 
fixed it to the reference value 
$\tan\beta=2$: as we will discuss 
in the following, larger/smaller values of $\tan\beta$ lead to 
smaller/larger effects in $\cB(\kpnn)$.\footnote{~Low values of $\tan\beta$ are 
strongly constrained by the experimental constraints  
on the Higgs sector, and in particular by the lower 
bounds on the lightest neutral Higgs~\cite{Higgs_ALEPH}.
We have explicitly checked that for $\tan\beta = 2$ 
(and even slightly below) we are compatible with these
experimental bounds. This happens 
mainly because we allow sizable trilinear soft-breaking terms in the up sector. }  
The impact of the phases of $\mu$ and $a_4$ has also been explicitly
studied, and found to be quite 
small (we will come back to this point later), so we considered only 
real parameters in the final scan. In any case, the different sign 
combinations of $\mu$ and $a_4$ allow to keep track of these phase effects.
Finally, in order to disentangle the various contributions, we have 
set $M_{H^+} > 1$~TeV, such that the charged-Higgs top-quark loops decouple. 
As chargino box effects can be safely neglected for a lightest slepton 
mass above 300 GeV, in practice we are left only with the 
chargino up-squark $Z$  penguin.

\begin{figure}[t]
\begin{center}
\vspace{-0.5 cm}
\includegraphics[scale=0.65]{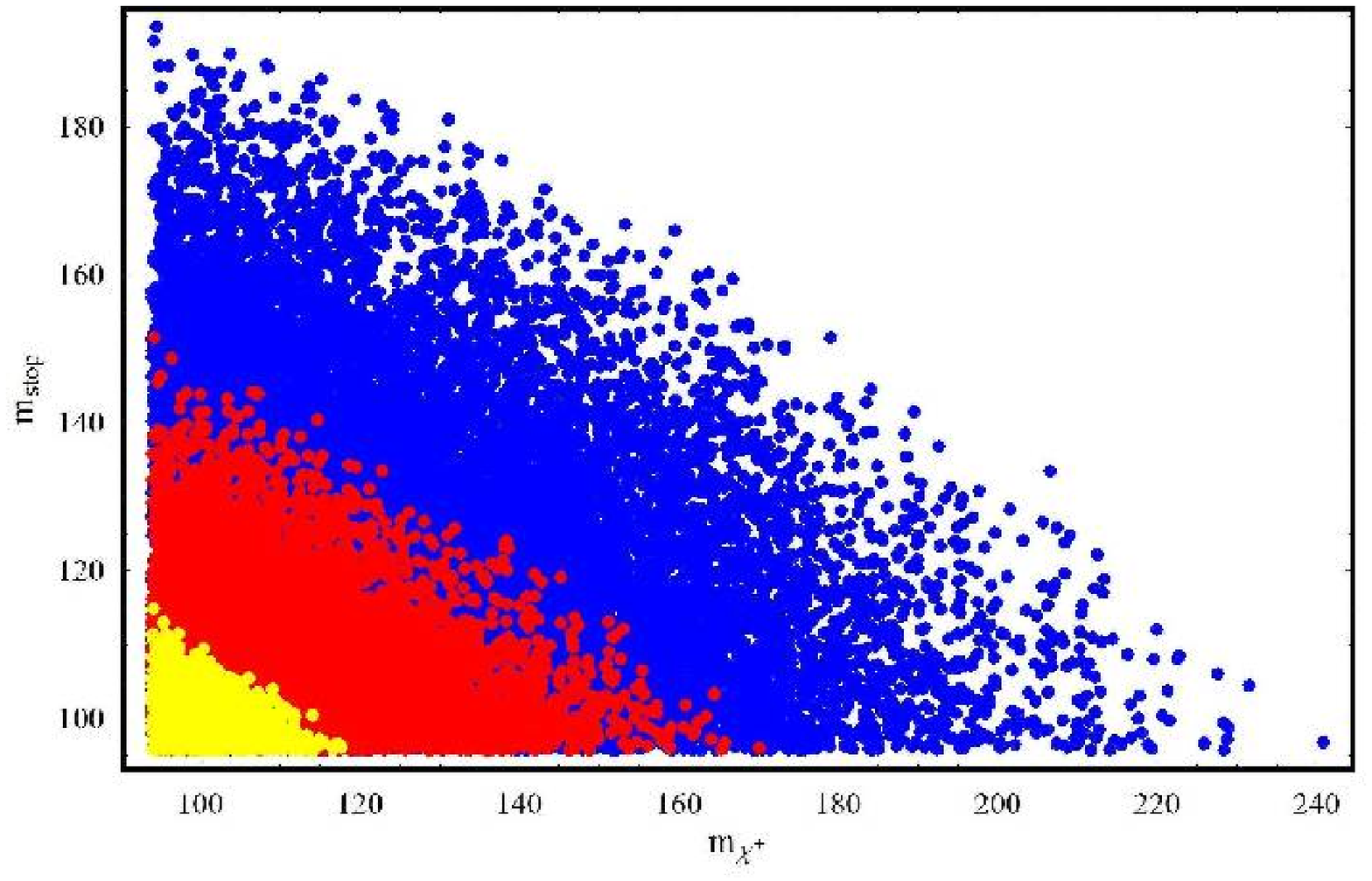}
\vspace{0.0 cm}
\caption{\label{fig:stopvschar} Regions in the
$m_\sttop$ -- $m_\charg$ plane (lightest stop and chargino masses)
allowing enhancements of $\cB(\kppn)$ 
of more than $11\%$ (yellow/light gray),  $8.5\%$ (red/medium gray) 
and $6\%$ (blue/dark gray) in the MFV scenario, for $\tan\beta=2$
and $M_{H^+} > 1$~TeV [the corresponding 
enhancements for $\cB(\klpn)$  are 15\%, 12.5\% and 10\%, 
respectively, see Eq.~(\ref{eq:RK})].}
\end{center}
\end{figure}

\begin{figure}[t]
\begin{center}
\vspace{-0.0 cm}
\includegraphics[scale=0.75]{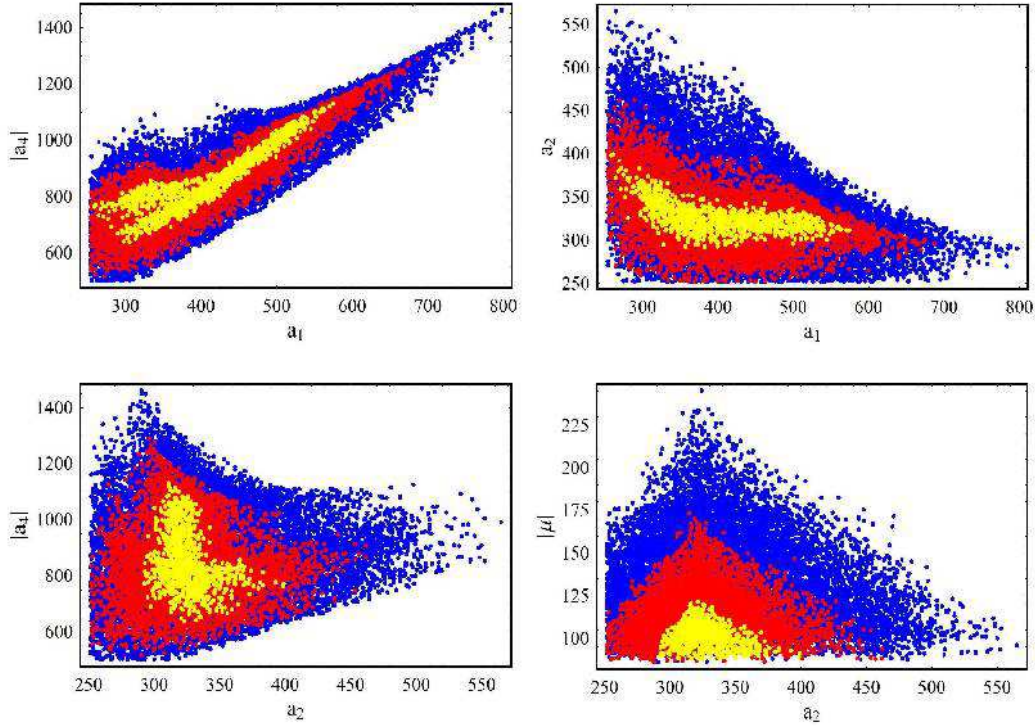}
\vspace{-0.0 cm}
\caption{\label{fig:a4vsa12} Correlations of the most significant 
MSSM-MFV parameters  corresponding to different enhancements 
of $\cB(\kppn)$ (identified by the color/gray-scale 
as in Figure~\ref{fig:stopvschar}). }
\end{center}
\end{figure}

The main results of the scan are illustrated in Figures~\ref{fig:stopvschar}
and~\ref{fig:a4vsa12}. As shown in Figure~\ref{fig:stopvschar},
the maximal enhancements of the branching ratios are closely
correlated to the minimal values for the lightest stop 
and chargino masses.\footnote{~The variation of density in these 
plots is not strictly correlated to the density of the underlying 
parameter space: by construction, we scan with more 
points the regions with larger effects.} This correlation 
is very useful to determine the flavour structure of the 
model. For instance, if  $\cB(\kppn)$ 
was found to be more than $10\%$ above the SM expectation 
and the lightest stop and chargino masses
were both found to be above $130$~GeV, with a charged Higgs mass above 
1 TeV, then one could exclude 
the MFV scenario. As expected, the MFV hypothesis predicts 
also a strict correlation between $\cB(\kppn)$ and $\cB(\klpn)$. We find in particular
\be
R(\kppn)  = (0.965\pm0.008) \times R(\klpn)~, 
\label{eq:RK}
\ee
\be
R(K\to f) = {\cB(K\to f)}/{\cB(K\to f)_{\rm SM}}~,
\ee
in the region of maximal enhancements (i.e.~$10\%$ to $16\%$ for the neutral mode).
In principle, the relation (\ref{eq:RK}) would allow the best test of the MFV hypothesis.
However, the experimental challenges of the   
$\klpn$ mode make the correlation between $\cB(\kppn)$,
$m_\sttop$ and $m_\charg$ outlined in Figure~\ref{fig:stopvschar}
a more useful test for the near future. 

In Figure~\ref{fig:a4vsa12}, we present a more detailed analysis 
of the parameter-space region with enhanced $\cB(\kpnn)$, showing 
the two-dimensional projections on the most significant planes. 
In this case is even more evident the key role of a precision 
measurement of $\cB(\kpnn)$ in selecting a well-defined region of
the model, or in constraining its structure.
As can be noted, an important role is played by the parameter $a_{4}$:
sizable enhancements of  $\cB(\kpnn)$ can occur only 
for large enough values of this parameter.
The reason of this effect can be traced back to the 
enhancement mechanism discussed in Ref.~\cite{CI}. 
Indeed, even within the MFV scenario one generates non-vanishing
left-right flavour-mixing terms 
in the squark basis of Refs.~\cite{BRS,CI}.
In particular, the double mass-insertion combination which controls 
possible enhancements in $\cB(\kpnn)$ \cite{CI,Jager} 
assumes the form\footnote{~We denote by $(\bar \delta_{LR}^{U})_{ij}$ the 
flavour-mixing couplings of Eq.~(\ref{eq:delta}) in the squark
 basis of Refs.~\cite{BRS,CI}.}
\begin{equation}
\left(\bar \delta_{RL}^{U}\right)_{32}^{\ast}\left( \bar \delta_{RL}^{U}\right)_{31}
\propto m_{t}^{2}V_{ts}^{\ast}V_{td}\left\vert a_{4}^{\ast}
-\mu\cot\beta\right\vert^{2} 
\label{eq:delta_MFV}
\end{equation}
and thus grows with $a_{4}$.
The expression (\ref{eq:delta_MFV})
also shows that: i) a possible CP-violating phase 
of $a_{4}$ has a negligible impact if $\mu$ is approximately real
(as implied by the stringent e.d.m.~bounds);
ii) maximal effects are obtained for $\mu$ and $a_{4}$ of opposite signs;
iii) small values of $\tan\beta$ enhance the magnitude of 
$(\delta_{RL}^{U})_{32}^{\ast}(\delta_{RL}^{U})_{31}$
 still further. 
We finally note that the contribution of 
$\mu$ is always subleading with respect to the one of $a_4$ 
(and thus the sensitivity to $\tan\beta$ is quite mild) 
since large values of $\mu$ lead to a strong suppression 
of the loop function (because of heavy higgsino masses).

\begin{figure}[t]
\begin{center}
\includegraphics[scale=0.5]{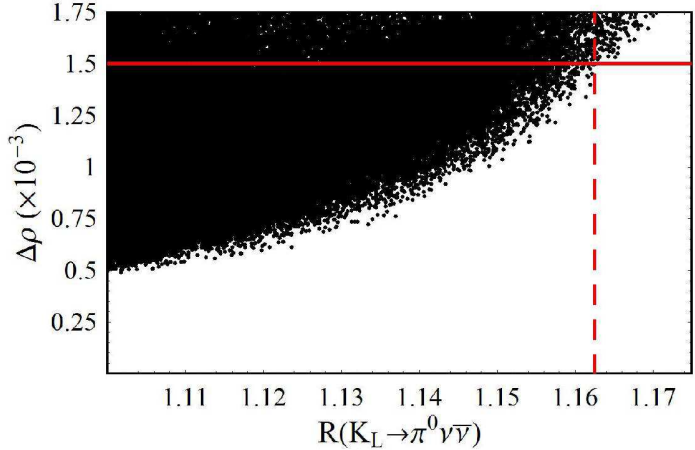}
\vspace{0.5 cm}
\caption{\label{fig:Deltarho}  Correlation 
between $\Delta\rho$ and $\cB(\klpn)$ in the MFV framework 
(with $\tan\beta=2$ and $M_{H^+} > 1$~TeV).}
\end{center}
\end{figure}

As far as the constraints from other observables are concerned, 
the MFV hypothesis automatically implies small non-standard 
effects in FCNC observables such as $\epsilon_K$ and 
$B\to X_s \gamma$. Moreover, these observables are sensitive 
to different combinations of free parameters with respect 
to the two $\cB(\kpnn)$. As a result, the requirement of 
consistency with existing FCNC data does not have a 
perceptible impact on Figures~\ref{fig:stopvschar} 
and~\ref{fig:a4vsa12}. As pointed out in Ref.~\cite{Kpnn_MFV1},
an important constraint is obtained from flavour-conserving 
electroweak observables and, in particular, from $\Delta\rho$. 
The correlation 
between $\Delta\rho$ and $\cB(\klpn)$ is illustrated in 
Figure~\ref{fig:Deltarho}. The effect for $\cB(\kppn)$
is almost identical, provided one scales the enhancement 
according to Eq.~(\ref{eq:RK}).

\subsection{Comparison with previous literature}

We conclude the MFV analysis with a comparison of our 
results with those obtained in a similar framework in 
Refs.~\cite{Kpnn_MFV1} and \cite{bobeth_pier}. 
First of all, we recall that for simplicity the numerical results
in Figures~\ref{fig:stopvschar} and~\ref{fig:a4vsa12} have been 
obtained in the limit where the charged-Higgs contribution 
can be neglected. This approximation can easily be removed:
charged-Higgs top-quark loops always induce a {\em constructive}
contribution to the $W$ function,
which depends only on $\tan\beta$ and $M_{H^+}$ (see e.g.~Ref.~\cite{CI}). 
For $\tan \beta=2$, the maximal effect amounts to 
an additional $\approx 5\%$ enhancement of the two $\cB(\kpnn)$
(whose limiting factor is the $M_{H^+} \gsim 300$~GeV 
bound derived from $B\to X_s\gamma$).
As a result, we conclude that in the MFV framework
\be
R(\kppn) \lsim 1.20~, \qquad  R(\klpn) \lsim 1.25~.
\label{eq:MFV_b}
\ee
This conclusion is in contradiction 
with respect to the claim of Ref.~\cite{Kpnn_MFV1}
that no sizable enhancement of the two  $\cB(\kpnn)$
is possible in the MFV framework. 
The origin of this difference is twofold:
\begin{itemize}
\item{}
We have considered a more 
general definition of the MFV framework
(the most general definition compatible 
with the renormalization group). Having a more 
general structure for the soft-breaking terms, 
within our scheme the effects in $\cB(\kpnn)$
are less severely constrained by the existing 
constraints on other observables.
\item{}
We have adopted a different strategy 
concerning the fit of the CKM matrix. In our analysis 
we have implicitly assumed that within the MFV framework 
the determination of the CKM matrix is not affected by 
the presence of new physics. Thanks to the recent precise 
results of $B$ factories, we now know that this assumption 
is an excellent approximation (see e.g.~\cite{Bona}). 
At the time of Ref.~\cite{Kpnn_MFV1}, the available 
experimental information was less precise and the authors 
decided to perform a non-standard CKM fit using observables 
sensitive to SUSY corrections. This fact  introduced 
spurious correlations between the genuine 
SUSY effects and indirect effects associated with the 
CKM determination. In particular, the predictions 
of the two $\cB(\kpnn)$ turned out to be suppressed because 
of smaller effective values of $\lambda_t$, and  not because 
of suppressed loop functions. 
\end{itemize}
In conclusion, we find that in the MSSM realization of the 
MFV hypothesis the two $\cB(\kpnn)$ can saturate the 
model-independent bounds of Ref.~\cite{bobeth_pier}.
This implies that a measurement of one of the two 
$\cB(\kpnn)$ consistent with Eq.~(\ref{eq:MFV_b})
does not allow to distinguish a generic MFV model from the MSSM. 
On the other hand, as illustrated by Figure~\ref{fig:stopvschar}, 
stringent tests of the model can be performed combining 
$\cB(\kpnn)$ and sparticle mass measurements.

\section{The general framework}
\label{sec:non_MFV}

As anticipated, within the MSSM there are in principle several new 
(non-Yukawa) sources of flavour-symmetry breaking.
Some of them are highly constrained by precise data on 
various rare processes. However, we are still far from being 
able to conclude that the MFV hypothesis, namely the absence 
of new sources of flavour-symmetry breaking, is the only viable option. 

Focusing the attention on the quark sector, we can 
identify five independent sources of flavour-symmetry breaking in 
the matrices $\MQ^2$,  $\MU^2$, $\MD^2$, $\AU$, and $\AD$.
Their non-trivial parts (terms not proportional to the identity matrix)
introduce breaking terms of the 
\be
G^{\rm quark}_{\rm flavour}=SU(3)_{Q_L} \times SU(3)_{U_R} \times SU(3)_{D_R}
\ee
subgroup of $SU(3)^5$ transforming respectively as 
\bea
&\MQ^2 \sim (8,1,1)~, \quad 
\MD^2 \sim (1,1,8)~,  \quad 
\AD \sim (\bar 3, 1, 3)~,& 
\label{eq:dtype} 
 \\
&\AU \sim (\bar 3, 3, 1)~,  \quad 
\MU^2 \sim (1,8,1)~.&
\label{eq:utype} 
\eea

In the literature, there exist several phenomenological 
analyses of such terms, typically expressed as upper 
bounds on the mass-insertion couplings $(\delta_{AB}^{U,D})_{ij}$,
defined as in Eq.~(\ref{eq:delta}). In particular, $\epsilon_K$ 
and $\Delta M_K$ imply stringent bounds on all the 1--2 
down-type mass insertions (limits in the $10^{-4}$--$10^{-3}$ range for squark masses below 500 GeV)
\cite{DMK}; bounds in the $10^{-2}$ range for all the 1--3 down-type couplings
follow from $\Delta M_{B_d}$ and $A_{\rm CP}(B_d\to J/\psi K)$ \cite{DBd};
bounds in the $10^{-2}$--$10^{-1}$ range are derived on  
the 2--3 down-type couplings from  $B \to X_s  \gamma$ \cite{DBs} 
(which strongly constrains LR terms) 
and recently also by $\Delta M_{B_s}$ \cite{DBs}
(which is also sensitive to LL and RR terms).
All these stringent phenomenological limits have been derived 
analysing the impact of gluino-mediated amplitudes. 
As a consequence, the constraints concern only down-type mass 
matrices, or the three flavour-symmetry breaking structures 
in Eq.~(\ref{eq:dtype}). The bounds on the  up-type 
soft-breaking terms, derived from  chargino amplitudes, 
are substantially weaker (see e.g.~Ref.\cite{Bsll} for a recent 
analysis in the 2--3 sector).

Interestingly, within the SM the only large breaking of the 
flavour symmetry appears in the up-sector, or in the up-type 
Yukawa coupling transforming as $(\bar 3, 3, 1)$.
It is therefore quite natural to conceive supersymmetric scenarios
where the three flavour-symmetry breaking structures in Eq.~(\ref{eq:dtype})
are very small (in agreement with observations) 
and sizable non-minimal breaking terms appear only in the 
up-type structures in Eq.~(\ref{eq:utype}), especially 
in the $(\bar 3, 3, 1)$ sector. 
As we shall show in the  rest of this section,
this scenario is perfectly compatible with all
existing constraints on $B$ and $K$ physics, and 
rare $K$ decays are the most useful tools to 
probe it in the future.

\subsection{Definition of the model}

We consider a non-minimal scenario where 
$\MQ^2$, $\MD^2$, and $\MU^2$ have an approximate MFV structure,
while $\AU$ contains sizable non-minimal flavour-breaking terms
of $\cO(\lambda)$, where  $\lambda \approx 0.22$ is the Cabibbo
angle. More precisely, we assume that in the super-CKM basis 
$\MuuLL^2$ and $\MuuRR^2$ have the form in Eq.~(\ref{eq:MFV_par}), while 
\be
\MuuRL^2 = \left[ \mathbf{A} -\cot\beta\mu^{\ast} \right]\mathbf{m}_{u}~,
\label{eq:MuLR_gen}
\ee
where 
\be
(\mathbf{A})_{33} \equiv A_0~, \qquad  \left|(\mathbf{A})_{i\not=j}\right| \leq \lambda A_0~.
\label{eq:Aij}
\ee 
This non-minimal structure is naturally 
consistent with  vacuum stability bounds \cite{vacuum}.
We note that structure of the type (\ref{eq:MuLR_gen}) also for 
$\MddRL^2$ (or  $\AD$),
with $\mathbf{m}_{u} \leftrightarrow \mathbf{m}_{d}$, 
is generally consistent with the existing bounds from 
gluino-mediated FCNC amplitudes (given the smallness of $\mathbf{m}_{d}$). 
However, in order to isolate the effects induced by $\AU$, 
in the following we concentrate on the case where $\AD$
is aligned to the corresponding Yukawa matrix.

In order to compare the sensitivity of various FCNC observables 
to the new sources of flavour-symmetry breaking, we fix the  
flavour-conserving parameters of the model to some reference values, 
and study the dependence of the observables on the $A_{ij}$ terms. 
This way we simulate somehow a post-LHC scenario, where most of the 
flavour-conserving parameters of the model are known, because
of the progress at the high-energy frontier, while precision
measurements of rare decays can be used to determine the 
flavour structure of the model. 

\begin{table}[t]
\begin{center}
\begin{tabular}{||clll||}
\hline
Charginos: &
$\mu = 500 \pm 10$ GeV  &  $M_2 = 300 \pm 10$ GeV  &  $\tan\beta =2$--$4$  \\
Up-squarks: &
$M_{\tilde u_R} = 600 \pm 20$ GeV
 & $M_{\tilde q_L} = 800 \pm 20$ GeV &  $A_0=1~{\rm TeV}$    \\
Other mass terms: & $M_1 = 500$~GeV
& \multicolumn{2}{l||}{  
$M_{\tilde d_R}=M_{\tilde l}=M_3=M_{H^+}=2$ {\rm TeV} }   \\
\hline
\end{tabular}
\end{center}
\caption{ \label{tab:ref_val} Basic choice of the flavour-conserving parameters 
used in Figures~\ref{fig:cern_1}--\ref{fig:fed_2}.}
\end{table}

\subsection{Numerical analysis}
 
The basic choice of the flavour-conserving parameters is shown in 
Table~\ref{tab:ref_val}. For simplicity, a high scale (2 TeV) has been chosen 
for all the parameters which play a minor role in chargino up-squark amplitudes.
The results depend very little from this choice, provided the minimum value 
of down-squark/slepton masses is above $500$ GeV. We have also assigned a 
small error to the key mass parameters of the chargino up-squark/sector, 
in order to investigate the sensitivity of flavour-changing observables 
to the precision on  mass terms. 
Given the smallness of the Yukawa couplings of the first 2 generations,
the only relevant  $A_{ij}$ terms are $A_{13}$, $A_{23}$, and $A_{33}\equiv A_0$.
As in the previous section, we have varied the free parameters 
of the model checking the consistency with direct experimental constraints 
on squark and gaugino masses and the existing 
constraints from rare processes (in particular the $B\to X_s \gamma$ one).


\begin{figure}[p]
\begin{center}
\includegraphics[scale=0.45,angle=-90]{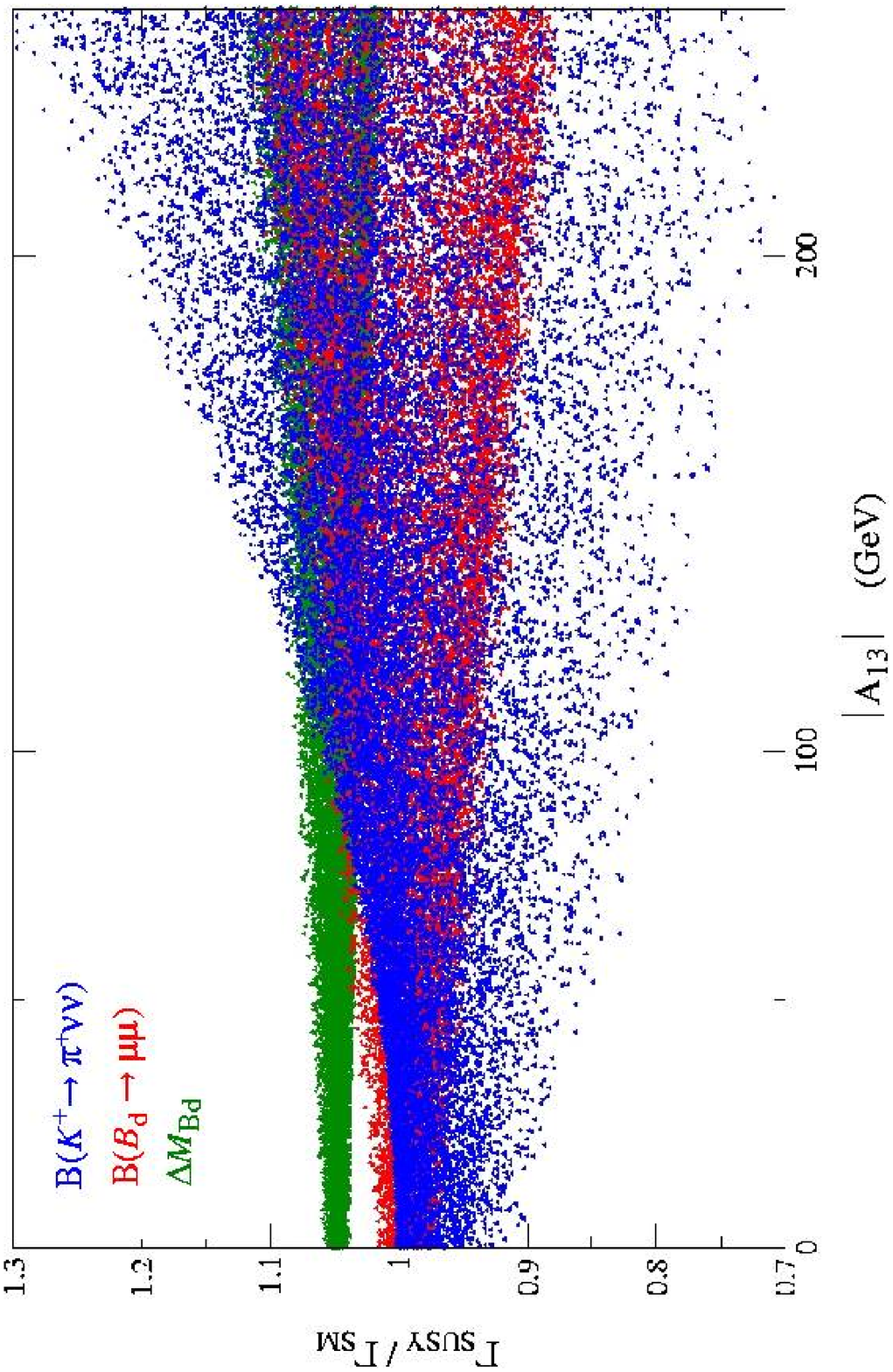} \\
\vspace{0.8 cm}
\includegraphics[scale=0.45,angle=-90]{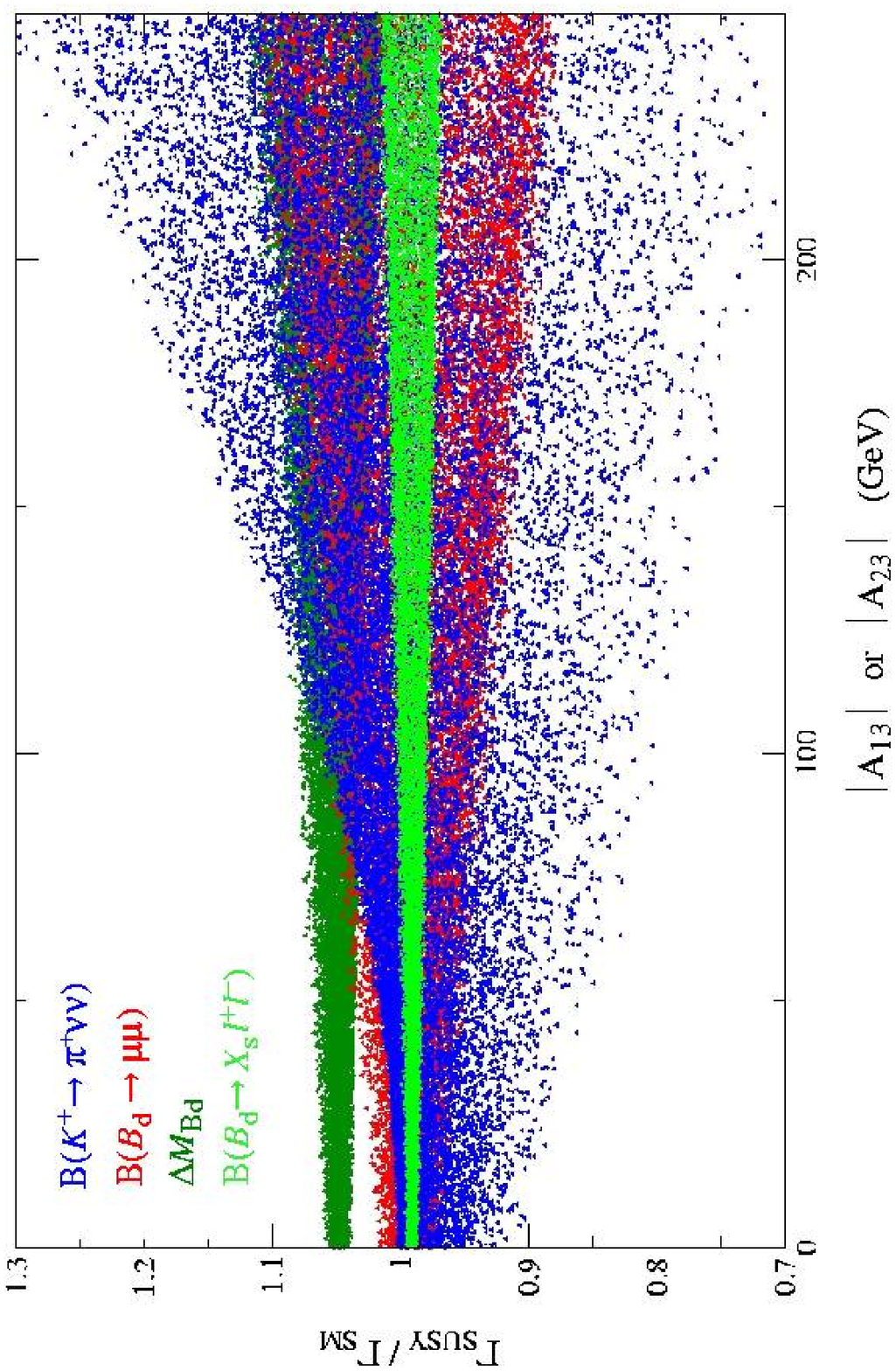} \\
\vspace{0.5 cm}
\caption{\label{fig:cern_1} Dependence of various FCNC 
observables (normalized to their SM value) on the up-type trilinear terms $A_{13}$ and $A_{23}$,
varied according to Eq.~(\ref{eq:Aij}). The flavour-conserving parameters
of the model are fixed as specified in Table~\ref{tab:ref_val}.
Upper plot: $\cB(\kppn)$ (blue/dark gray), $\cB(B_d\to\mu^+\mu^-)$ (red/gray lower-region), 
$\Delta M_{B_d}$ (green/gray upper-region) as a function of $A_{13}$.
Lower plot: same observables as in the upper plot, with the superposition 
of $\cB(B_d\to X_s\ell^+\ell^-)$ (light green/light gray)
plotted as a function of $A_{23}$ (instead of $A_{13}$). }
\end{center}
\end{figure}

\begin{figure}[t]
\begin{center}
\includegraphics[scale=0.55,angle=-90]{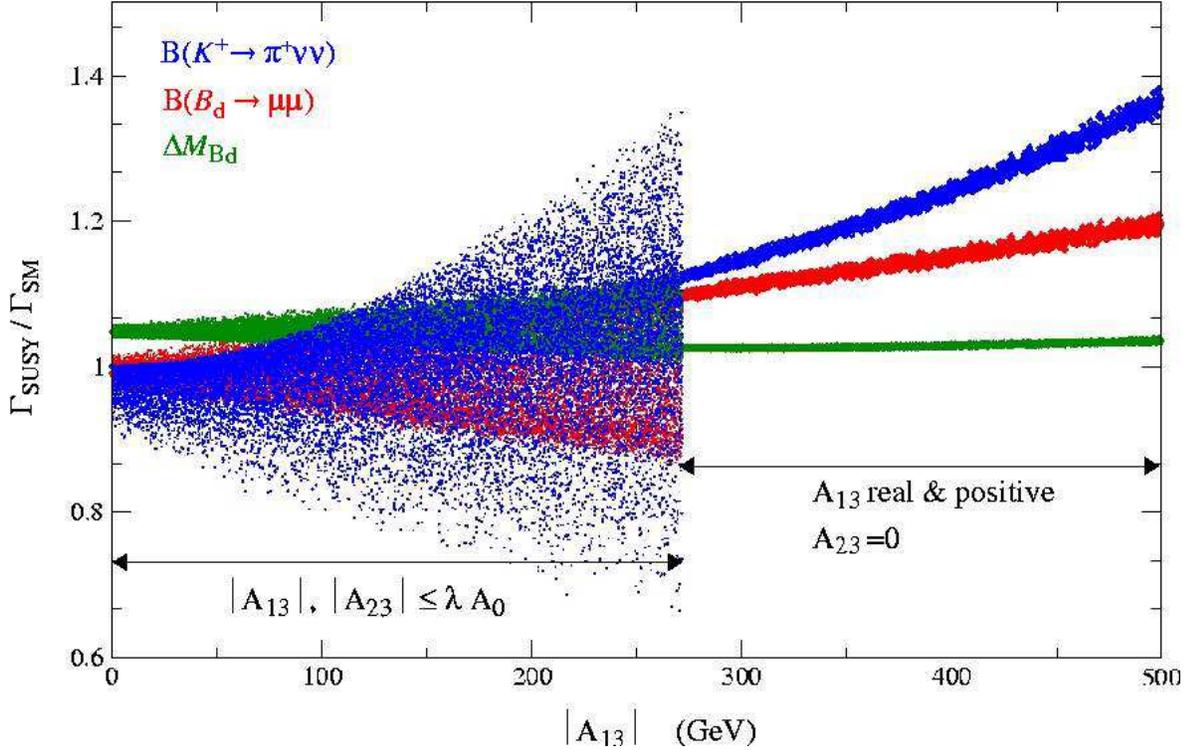}\\
\vspace{1.0 cm}
\caption{\label{fig:cern_1b} 
Anatomy of the SUSY contributions to $\cB(\kppn)$, 
$\cB(B_d\to\mu^+\mu^-)$ and $\Delta M_{B_d}$ changing the range of 
variability of $A_{13}$ and $A_{23}$. Notations and other conventions 
as in Figure~\ref{fig:cern_1}. }
\end{center}
\end{figure}

\begin{figure}[p]
\begin{center}
\includegraphics[scale=0.43,angle=-90]{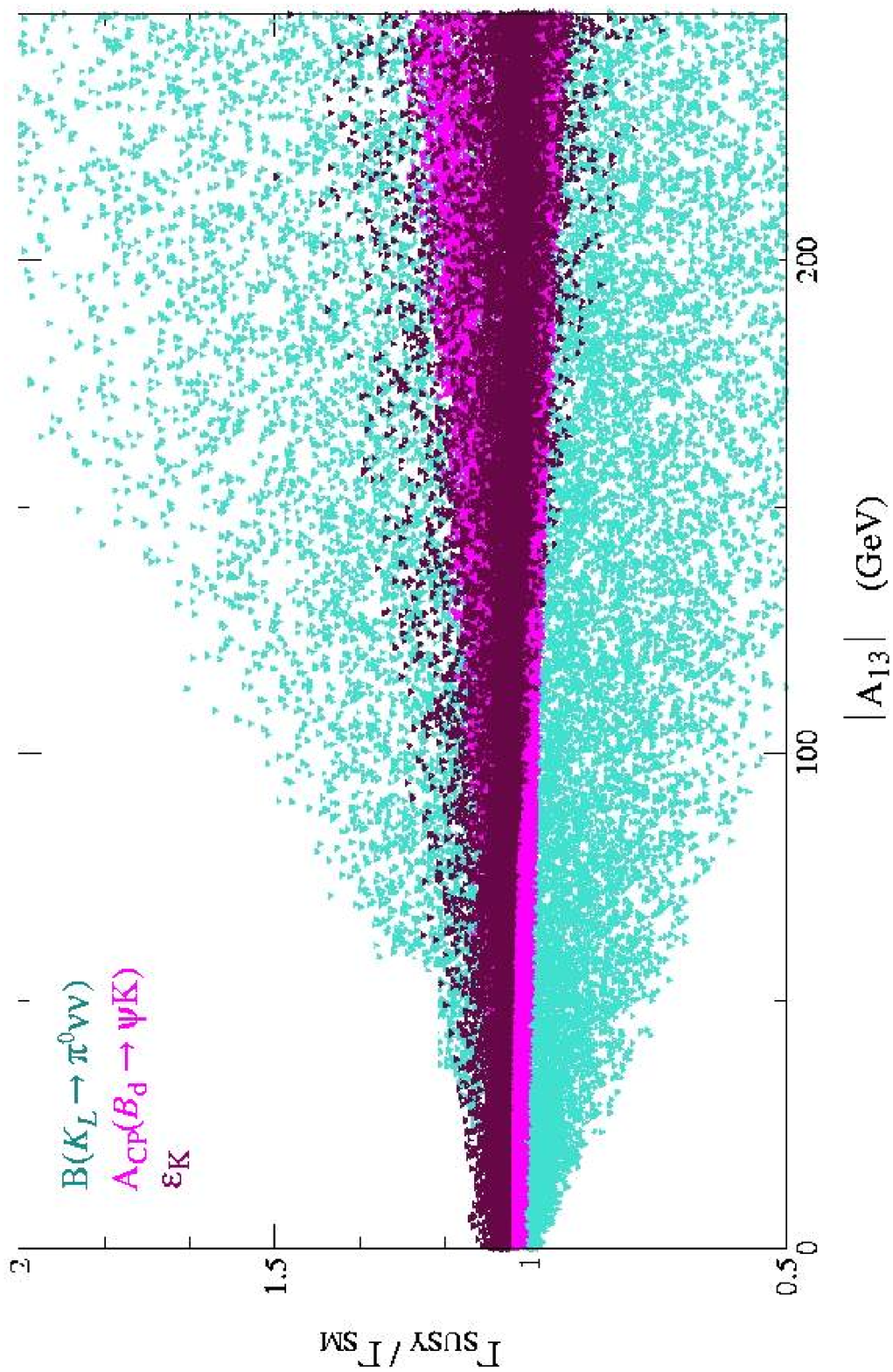} \\
\vspace{1.0 cm}
\includegraphics[scale=0.43,angle=-90]{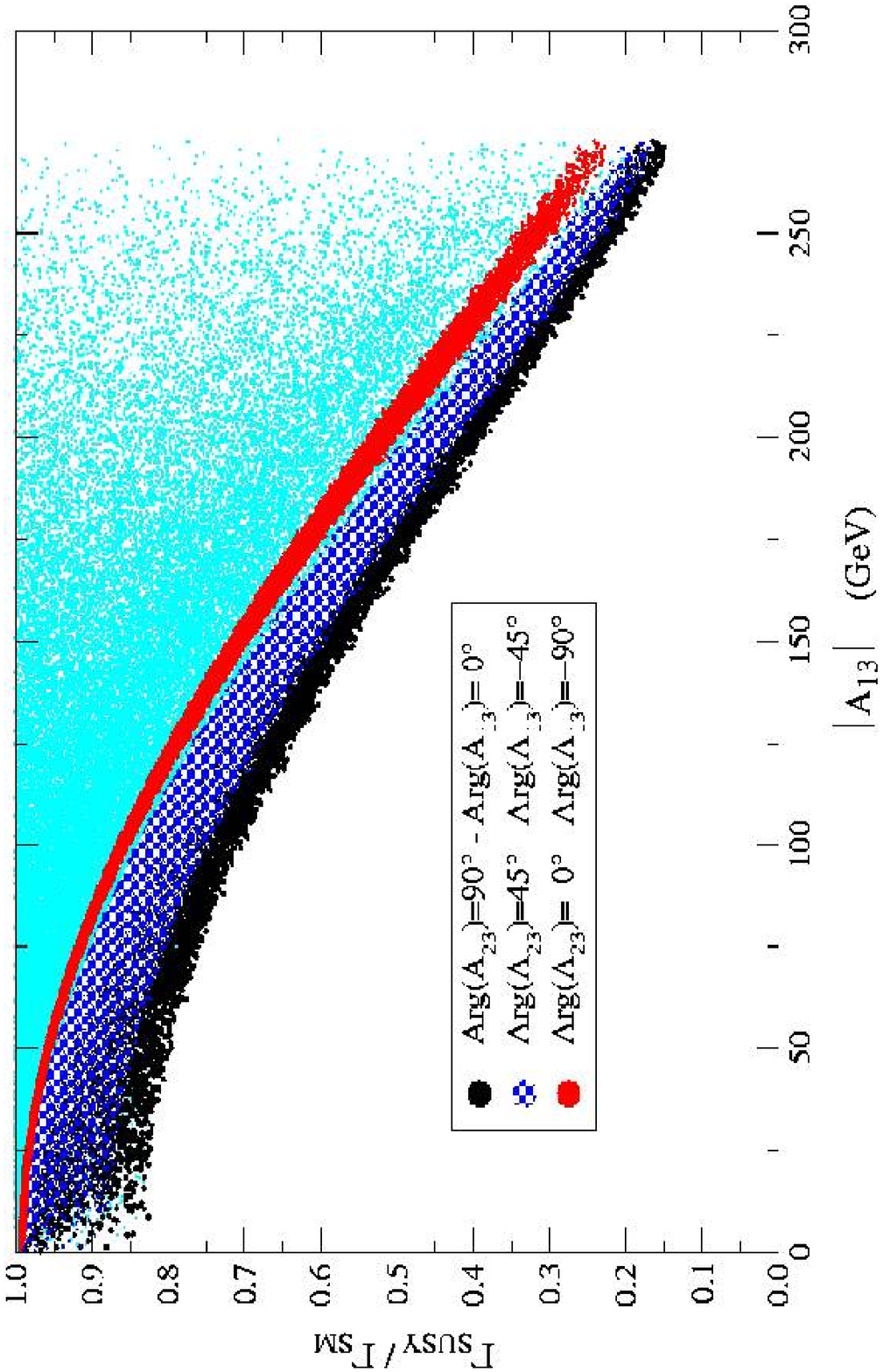} \\
\vspace{1.0 cm}
\caption{\label{fig:cern_2} 
Impact of chargino-mediated amplitudes on CP-violating observables. 
Upper plot: comparison of $\epsilon_K$ (bordeaux/dark gray), $A_{\rm CP}(B_d\to J/\psi K)$
(violet/gray) and $\cB(\klpn)$ (light blue/light gray).
Lower plot: dependence of the non-standard contributions 
to $\cB(\klpn)$ on the CP-violating phases of $A_{13}$ and $A_{23}$ 
in the region of destructive interference between SUSY and SM contributions
(light-blue/light-gray dots correspond to no constraints on the phases).
The trilinear terms are varied according to Eq.~(\ref{eq:Aij}); all the 
other supersymmetric parameters are fixed as in Table~\ref{tab:ref_val}.}
\end{center}  
\end{figure}

\begin{figure}[t]
\begin{center}
\includegraphics[scale=0.55,angle=-90]{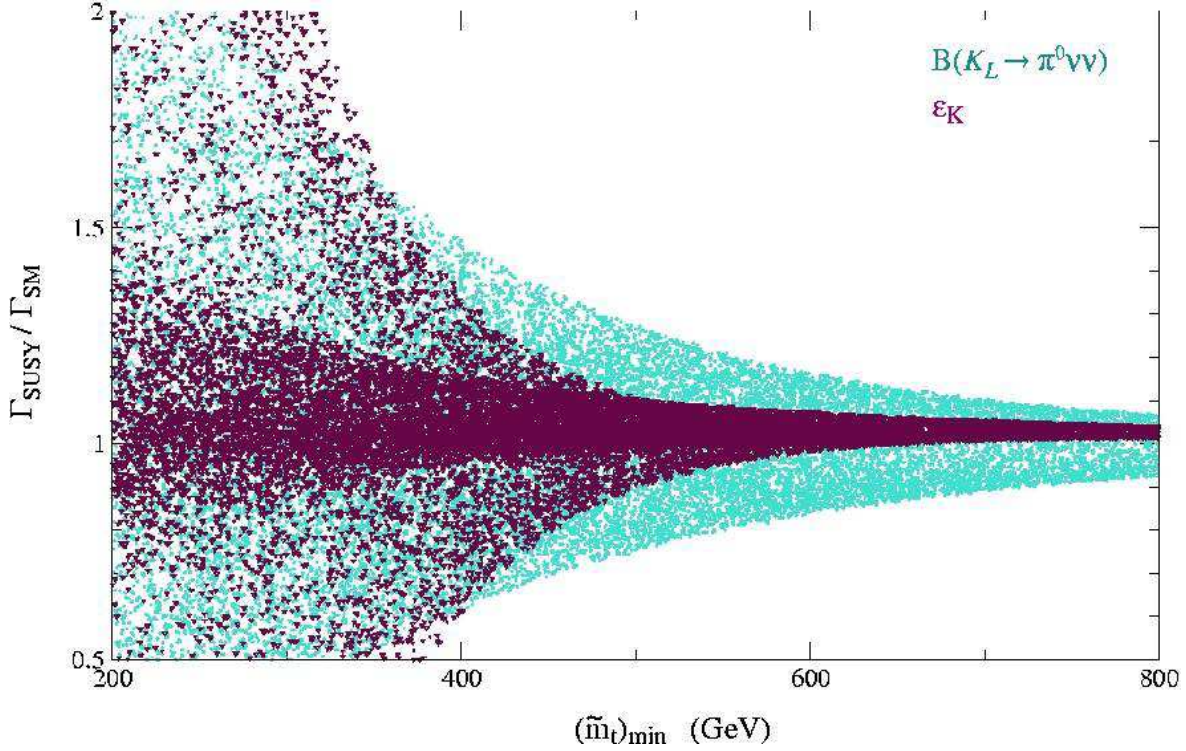} \\
\vspace{1.0 cm}
\caption{\label{fig:cern_3} 
Comparison of the decoupling of non-standard contributions to 
$\epsilon_K$ (bordeaux/dark gray) and $\cB(\klpn)$ (light blue/light gray)
in the limit of heavy supersymmetric particles. The scatter plots are obtained 
varying $A_{13}$ and $A_{23}$ according to Eq.~(\ref{eq:Aij}), 
$M_{\tilde u_R}$ in the interval 200--1000 GeV, and fixing all the 
other mass parameters as in Table~\ref{tab:ref_val}. The 
horizontal axis denotes the lightest up-type squark mass. }
\end{center}
\end{figure}

\begin{figure}[t]
\begin{center}
\includegraphics[scale=0.5,angle=-90]{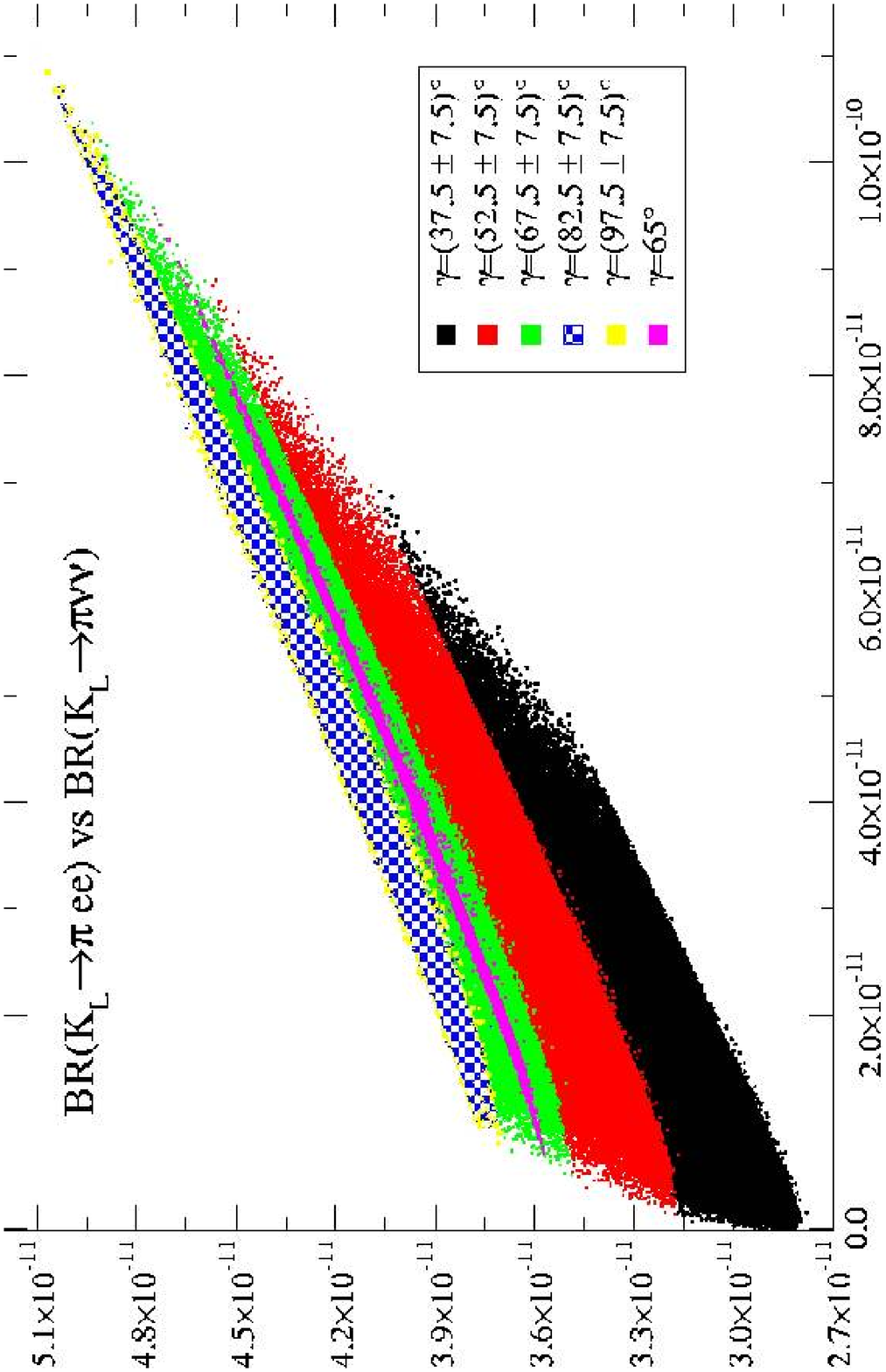} \\
\vspace{1.0 cm}
\caption{\label{fig:fed_2} 
Correlation between  $\cB(\klpn)$ (horizontal axis) and $\cB(\kLpee)$ 
(vertical axis) in the MSSM scenario with non-minimal   $A_{13}$ and $A_{23}$,
varied according to Eq.~(\ref{eq:Aij}) (other parameters are fixed as in Table~\ref{tab:ref_val}).
Different colors (gray scales) correspond to different values of the CKM 
phase $\gamma$. }
\end{center}
\end{figure}


The numerical results thus obtained for the two $K\to \pi\nu\bar\nu$
modes, in comparison with $B_d\to X_s\ell^+\ell^-$, $B_d\to\mu^+\mu^-$,
$\Delta M_{B_d}$, and the CP-violating observables 
$\epsilon_K$ and $A_{\rm CP}(B_d\to J/\psi K)$, are 
shown in Figures~\ref{fig:cern_1}--\ref{fig:fed_2}.
The main features resulting from this numerical study
can be summarized as follows:
\begin{itemize}

\item 
The non-standard effects induced by these chargino-mediated 
amplitudes, in the presence of non-MFV up-type $A$ terms,
are maximal in the two $K\to\pi\nu\bar\nu$ decays. 
The dominance of $K\to\pi\nu\bar\nu$ holds in comparison with other 
$K$- and $B$-physics FCNC amplitudes, both in CP-conserving 
(Figure~\ref{fig:cern_1}) and in CP-violating observables 
(Figure~\ref{fig:cern_2}). 

Note that the non-standard effect in $\cB(\kpnn)$
is not necessarily an enhancement with respect to the SM. For instance, 
for $A_{13}\approx 0$ one has  $\cB(\kppn) \lsim \cB(\kppn)_{\rm SM}$,
while in the same region of parameter space $\Delta M_{B_d}$ 
receives a $\approx 4\%$  positive correction (see Figure~\ref{fig:cern_1}).
However, the important feature emerging from our analysis 
is that $\cB(\kppn)$ has a much stronger 
dependence on $A_{13}$ with respect to $\Delta M_{B_d}$
(and the other  $B$-physics observables).
\item
As can be noted from Figures~\ref{fig:cern_1}--\ref{fig:cern_2}, 
despite the fact that the flavour conserving parameters of the model are 
completely fixed, the predictions for the FCNC observables 
are still affected by a  large uncertainty. This is due to the 
free parameters hidden in phases and moduli of $A_{13}$ and $A_{23}$
(see Figure~\ref{fig:cern_1b}). 
For this reason, a complete determination of the model 
requires several precision measurements in the FCNC sector.
Ideally, the two $K\to\pi\nu\bar\nu$ rates and  
two clean FCNC observables in the $B$ system (such as 
the rates of the two $B_{s,d} \to\mu^+\mu^-$ modes).

\item 
As can be seen in Figure~\ref{fig:cern_1b}, 
even setting $A_{23}=0$, the dependence of $\cB(\kpnn)$ on 
$A_{13}$ is quadratic, contrary to the approximate linear 
dependence of the other FCNC observables. This quadratic 
dependence, which is the main reason of the dominance of 
non-MFV chargino-mediated amplitudes in $K\to\pi\nu\bar\nu$,
can be explained in terms of the double-mass insertion 
mechanism of Ref.~\cite{CI}. 

To better understand this effect, we denote by $A_{ij}$ 
($\bar A_{ij}$) the $A$ terms in the super-CKM basis
(the squark basis of Refs.~\cite{BRS,CI}). It is easy 
to realize that  if $A_{23}=0$ and  $A_{13}\not=0$ 
in the super-CKM basis, then both ${\bar A}_{13}$
and ${\bar A}_{23}$ are not vanishing because of the non-diagonal 
entries of the CKM matrix. In particular, 
\be
{\bar A}_{23} = V_{2j} A_{j3} \approx V_{21} A_{13}~.
\ee
A similar mechanism occurs if  $A_{13}=0$ and $A_{23}\not = 0$. 
Thus, barring fine-tuned scenarios, there is always a 
sizable effective double mass-insertion coupling, given by 
\begin{equation}
\left(\bar \delta_{RL}^{U}\right)_{32}^{\ast}\left( \bar \delta_{RL}^{U}\right)_{31}
\propto  {\bar A}_{23} {\bar A}_{13}^* \approx  
V_{12} |A_{23}|^2 +  V_{21} |A_{13}|^2 + A_{23} A_{13}^*
\label{eq:nnn}
\end{equation}
Since the double mass-insertion enhancement is effective only in the 
kaon system (because of its double-CKM suppression \cite{CI}), 
rare $K$ decays are naturally the best probe of 
any non-MFV structure in $\AU$. 

A further check of the dominance of the double mass insertion
mechanism is shown in the lower plot of Figure~\ref{fig:cern_2}, 
for the CP-violating channel $\klpn$. Here both $A_{2 3}$ and $A_{13}$
are non vanishing, and regions corresponding to specific 
choices of their CP-violating phases are outlined with different 
colors (gray scales). As can be noted, to a good approximation  
the non-standard effect depends only on their relative phase 
and --consistently with Eq.~(\ref{eq:nnn})-- 
is maximal when  $A_{23} A_{13}^*$ is purely imaginary.

\item   
At first sight, it is quite surprising that the non-MFV
left-right mixing terms in Eq.~(\ref{eq:Aij})
are not excluded by the precise data on $\cB(B\to X_s \gamma)$
and have a marginal impact also on 
 $\cB(B\to X_s \ell^+\ell^-)$ (see Figure~\ref{fig:cern_1}).
However, this fact can easily be understood by noting that 
a non-vanishing $b \to s \gamma$ amplitude (generated by effective 
operators of the type $\bar b_{R} \sigma_{\mu\nu} s_{L} F^{\mu\nu}$
or $\bar b_{L} \sigma_{\mu\nu} s_{R} F^{\mu\nu}$)
requires:
\begin{enumerate}
\item[i.]  odd number of chirality flips in the down sector;
\item[ii.] odd total number of chirality flips summing up, down and 
flavour-independent terms.
\end{enumerate}
 This implies 
that the up-type left-right mixing terms 
in Eq.~(\ref{eq:Aij}) can contribute 
to the $b \to s \gamma$ transition only via amplitudes 
which have at least three chirality flips (the up-type
trilinear term, one left-right mixing in the down sector, 
and a third $SU(2)_L$ breaking term in order to recover the 
total helicity-violating structure). Since each chirality flip 
is associated with an insertion of the SM Higgs vev, 
this structure implies a strong suppression.

A phenomenological check of this statement is provided 
by the loose  bounds on $(\delta_{RL}^{U})_{32}$
extracted from $\cB(B\to X_s \gamma)$ (see e.g.~Refs.~\cite{Bsll,Misiak}).

\item 
An important property of the chargino-mediated non-MFV contributions 
is their slow decoupling in the limit of heavy 
superpartners in $\kpnn$ decays 
 ($\sim m^{-2}_{\rm SUSY}$), compared to the 
fast decoupling  in $\Delta F=2$ observables ($\sim m^{-4}_{\rm SUSY}$).
This property, which has been discussed in detail in  Ref.~\cite{Kpnn_SUSY1}, 
is clearly illustrated in Figure~\ref{fig:cern_3}
in the case of CP-violating observables
(a completely similar scenario holds in the CP-conserving case).
This property provides another natural explanation of why we 
can still hope to observe sizable deviations from the SM 
in rare $K$ decays, despite the absence of non-standard 
effects in $\Delta M_{B_{d,s}}$, $A_{\rm CP}(B_d \to J/\psi K)$,
and $\epsilon_K$. 

\item 
A well-defined prediction of this non-MFV scenario 
is a strict correlation between the non-standard 
contributions to 
$\klpn$, $\kLpee$, and $\kLpmm$. As shown in Figure~\ref{fig:fed_2}, 
the relative size of the effect is larger 
in the $\klpn$ case; however, observable deviations 
from the SM are expected also in the two 
$\kLpll$ modes. This correlation is a 
consequence of the $Z\bar s d$ effective vertex
common to the three decay modes, which  
encodes the dominant non-SM contributions
in the limit of heavy sleptons.
The detailed structure of the correlation is almost 
independent of the flavour-conserving parameters 
of the MSSM, for sufficiently heavy sleptons,
but has a significant dependence of the CKM
phase $\gamma$ (as shown in Figure~\ref{fig:fed_2}).
 
If at least two of the rare $K_L$ channels 
could be measured with good precision, their
correlation within this non-MFV scenario
would provide a striking signature of the model.

\end{itemize}

\section{Conclusions}
The determination of the flavour structure of the MSSM 
--as well as of any TeV extension of the SM--
is one of the components of the so-called {\em inverse problem}~\cite{Nima},
which hopefully we will face soon with the start of the LHC program.
While the direct exploration of the TeV scale will 
reveal some of the features of the new-physics scenario, 
the LHC program alone is unlikely to completely 
determine the structure of the new underlying theory.
This statement is particularly true in the case 
of the flavour structure of the theory, whose model-independent determination 
requires new high-precision measurements also at low energies.
 
In this paper we have provided a quantitative 
illustration of this problem, analysing the impact 
of the two $K\rightarrow\pi\nu\overline{\nu}$ modes 
in determining the flavour structure of the MSSM. 
We have analysed the expectations for the branching ratios 
of these modes in two representative classes of the MSSM
(as far as the flavour structure is concerned): 
the Minimal Flavour Violation scenario \cite{MFV}
and a scenario with  $\AU$ terms not aligned with 
the corresponding Yukawa coupling.
In both these frameworks precise measurements 
of the two $\cB(K \to \pi \nu\overline{\nu})$
turn out to be key ingredients to determine 
the structure of the model. 

Within the MFV scenario, the deviations from the SM 
in the two $\cB(K \to \pi \nu\overline{\nu})$ are 
naturally small (typically within $10\%$), but could
saturate the model-independent bounds of Ref.~\cite{bobeth_pier}. 
We have outlined some clean correlations between the possible 
non-standard effects in the two $\cB(K \to \pi \nu\overline{\nu})$ 
and the values of stop and chargino masses which provide 
a distinctive signature of this scenario, or an efficient 
method to test the implementation of the MFV hypothesis 
in the MSSM.

The situation is certainly more interesting in the 
case of non-MFV  up-type trilinear terms. Within this 
scenario, which is well motivated and perfectly compatible 
with all existing constraints from $B$ and $K$ physics,
the two $\cB(K \to \pi \nu\overline{\nu})$ could receive 
$\cO(1)$ corrections with respect to the SM.
We have indeed demonstrated that these rare $K$ decays, 
together with the two $K_L \to \pi^0 \ell^+\ell^-$ modes, 
represent the most sensitive probe of any misalignement 
between $\AU$ and the corresponding Yukawa coupling.
The precise measurement of these rare decays 
is therefore a necessary ingredient for a
model-independent reconstruction of the flavour 
structure of the MSSM soft-breaking terms.

\subsection*{Acknowledgements} 
We thank Andrea Brignole and Simone Pacetti for useful discussions.
This work is partially supported by IHP-RTN, 
EC contract No.\ HPRN-CT-2002-00311 (EURIDICE).
The work of C.S.~is also supported by the Schweizerischer Nationalfonds;
S.T.~acknowledges the support of the  
DFG grant No.~NI 1105/1-1 and DFG within SFB/TR-9 (Computational Particle Physics).

\end{document}